\documentclass[a4paper,11pt]{article}
\usepackage{pos}

\title{Forward silicon tracking detector developments for the future Electron-Ion Collider}

\author*[a]{Xuan Li}
\author[a]{Melynda Brooks}
\author[a]{Matt Durham}
\author[a]{Ming Liu}
\author[a]{Yasser Corrales Morales}
\author[a]{Kei Nagai}
\author[a]{Anton Navazo}
\author[a]{Christopher Prokop}
\author[a]{Eric Renner}
\author[a]{Walter Sondheim}
\author[a]{Cesar da Silva}
\affiliation[a]{Los Alamos National Laboratory,\\
  Los Alamos, NM, USA} 

\emailAdd{xuanli@lanl.gov}

\abstract{The future Electron-Ion Collider (EIC) will utilize a series of high-luminosity high-energy electron+proton ($e+p$) and electron+nucleus ($e+A$) collisions to explore the inner structure of nucleon and nucleus and the matter formation process. Heavy flavor hadron and jet measurements at the EIC will play an essential role in determining the nucleon/nucleus parton distribution function and heavy quark hadronization process in the not well constrained kinematic region. A high granularity and low material budget forward silicon tracker will enable precise forward heavy flavor measurements at the EIC, which have enhanced sensitivities to access these kinematic extremes. A Forward Silicon Tracker (FST) detector is under detector design and R$\&$D for the EIC. Two advanced silicon technologies, the Depleted Monolithic Active Pixel Sensor (DMAPS) and the AC coupled Low Gain Avalanche Diode (AC-LGAD), which can provide fine spatial and timing resolutions, have been considered as candidates for the EIC silicon tracking detector. Progresses and results about the FST conceptual design and ongoing DMAPS and LGAD detector R$\&$D will be presented. The path forward towards an integrated EIC detector will be discussed as well.}

\FullConference{%
  41st International Conference on High Energy physics - ICHEP2022\\
  6-13 July, 2022\\
  Bologna, Italy
}

\begin{document}
\maketitle

\section{Introduction}
The next generation Quantum Chromodynamics (QCD) facility: the Electron-Ion Collider (EIC) will be built at the Brookhaven National Laboratory to study the nucleon/nucleus inner structure, solve the proton spin puzzle and explore how normal matter is formed from fundamental quarks and gluon. The future EIC, which will operate high energy high luminosity electron+proton ($e+p$) and electron+nucleus ($e+A$) collisions, has received the CD0 and CD1 approval from US Department of Energy. To realize various EIC physics measurements \cite{eic_YR}, a high granularity multiple layer detector, which can provide precise measurements of primary and displaced vertice, track reconstruction, particle identification and energy information, is required. Heavy flavor hadron and jet measurements at the EIC can significantly improve existing understanding of the nuclear Parton Distribution Functions (nPDFs) and the heavy quark hadronization process in vacuum and nuclear medium. A Forward Silicon Tracker (FST) detector plays an important role in reconstructing forward heavy flavor products in the asymmetric $e+p$ and $e+A$ collisions at the EIC. The conceptual design of the proposed FST for the future EIC, the forward tracking performance and results of the related advanced silicon sensor R$\&$D will be discussed.

\section{Forward Silicon Tracker Detector Conceptual Design}
\begin{figure}[!ht]
\centering
\includegraphics[width=0.6\textwidth]{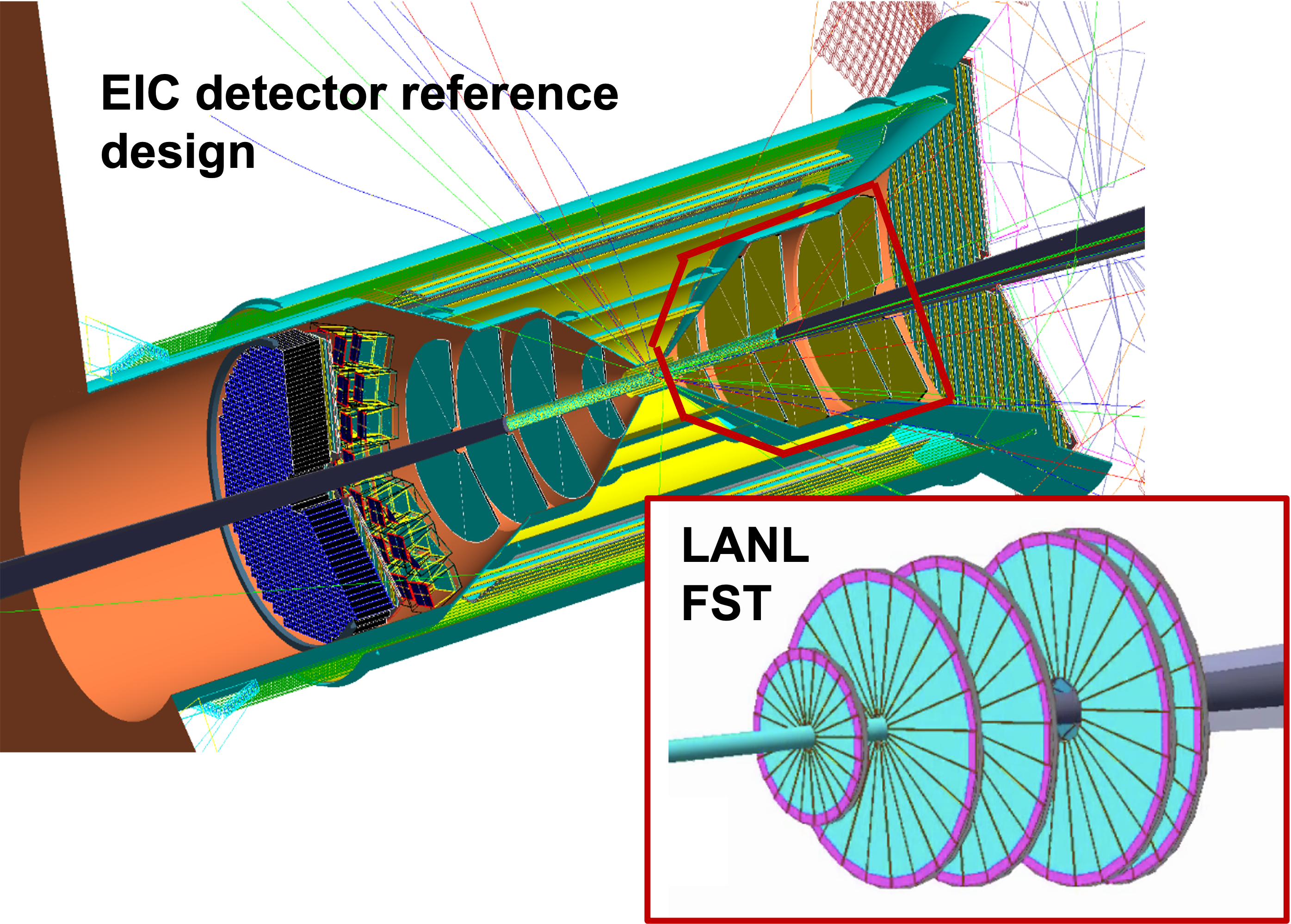}
\caption{The Forward Siliocn Tracker (FST) \cite{lanl_eic, lanl_fst} geometry has been implemented in GEANT4 simulation as highlighted inside the red box. The FST design, which consists of detailed layout of the segmented silicon sensor wedges, readout modules and basic support structure, has been included in the EIC detector reference design \cite{ecce}.}
\label{fig:fst_design}
\end{figure}

The future EIC will operate a series of $e+p$ and $e+A$ collisions with the instant luminosity of $10^{33-34} cm^{-2}s^{-1}$ and center of mass energies from 29~GeV to 141~GeV \cite{eic_YR}. The electron beam energy is from 5~GeV to 18~GeV and the proton/nucleus beam energy is from 41~GeV to 275~GeV. The bunch crossing rate is around 10~ns. Particles produced in the asymmetric $e+p$ and $e+A$ collisions at the EIC have higher cross sections in the forward pseudorapidity region (nucleon/nucleus beam going direction) than that in the backward pseudorapidity region (electron beam going direction). This feature requires that the EIC detector to provide a wide kinematic coverage for forward particle measurements \cite{eic_YR}. A Forward Silicon Tracker (FST) detector, which consists of 5 silicon disks based on the Monolithic Active Pixel Sensor (MAPS) \cite{maps_alpide} technology, has been proposed by the Los Alamos National Laboratory (LANL) EIC team \cite{lanl_eic, lanl_fst}. The proposed FST can achieve fine spatial resolution and low material budgets in the pseudorapidity region of $1.2 < \eta < 3.5$. The FST conceptual design has been included in the selected EIC detector reference design \cite{ecce}. Figure~\ref{fig:fst_design} shows the geometry of the EIC tracking detector reference design implemented in GEANT4 \cite{geant4} simulation, in which the FST has been included.

\begin{figure}[!ht]
\centering
\includegraphics[width=0.75\textwidth]{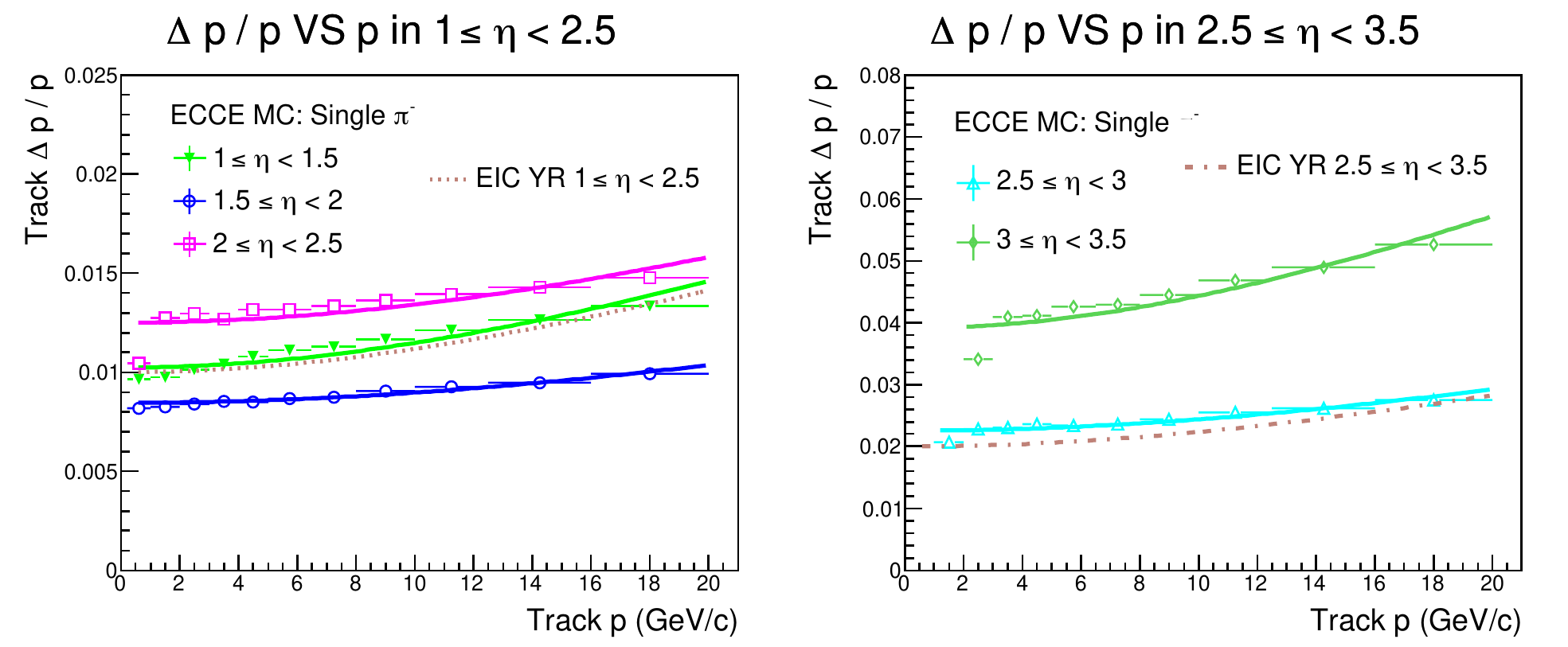}
\includegraphics[width=0.75\textwidth]{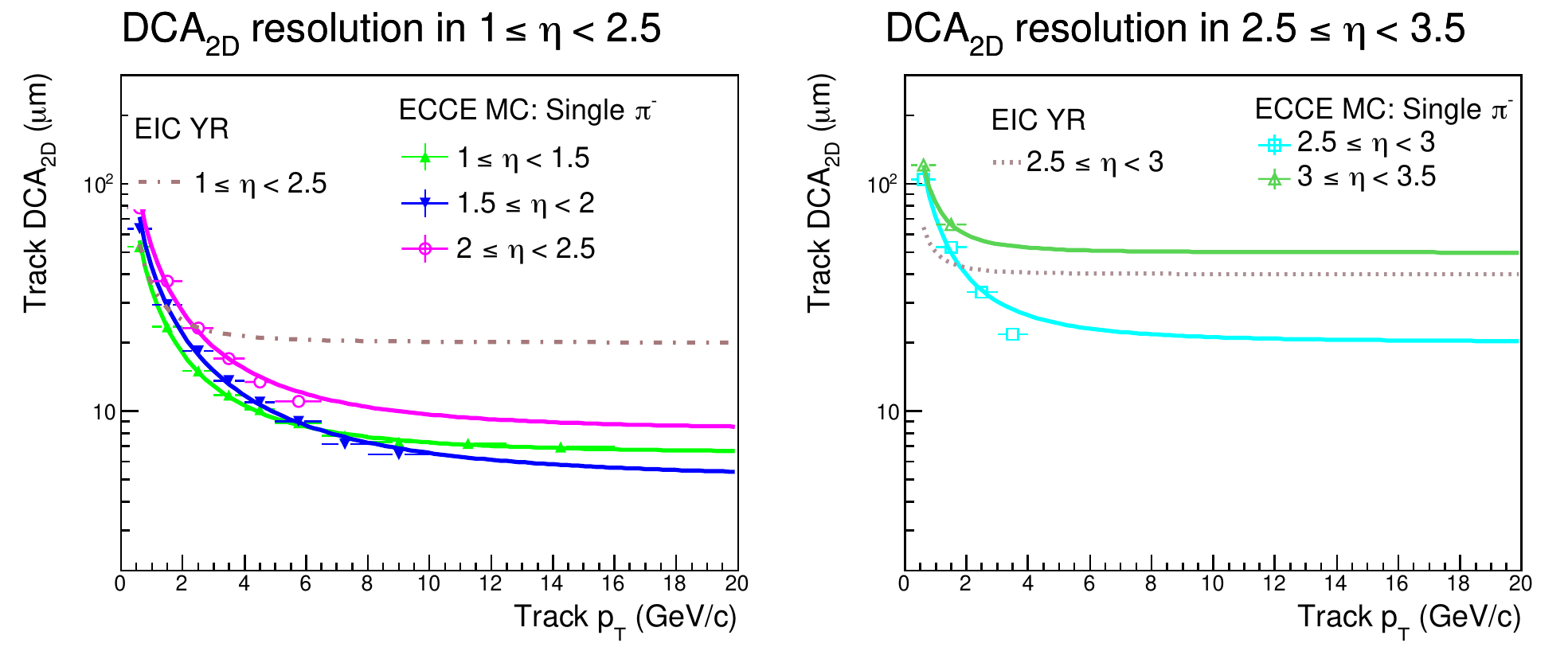}
\caption{The tracking performance of the EIC detector reference design in the psedurorapidity region of $1.0 \leq \eta \leq 3.5$. The track momentum dependent momentum resolution (top panel) and the track transverse momentum dependent transverse Distance of Closest Approach ($DCA_{2D}$) resolution in different pseudorapidity bins are compared with the EIC yellow report requirements \cite{eic_YR}. These performance results are evaluated with the 1.4~T Babar magnet.}
\label{fig:fst_perform}
\end{figure}

The FST plays an essential role in track reconstruction in the forward peudorapidity region for the EIC project detector. The track performance of the EIC detector reference design \cite{ecce} has been studied in simulation. The track momentum dependent momentum resolutions and the transverse momentum dependent transverse Distance of Closet Approach ($\text{DCA}_{\text{2D}}$) resolutions in the pseudorapidity region of $1.0 \leq \eta \leq 3.5$ with the Babar magnet are shown in the top and bottom panels of Figure~\ref{fig:fst_perform} respectively. These tracking resolutions are compared to the EIC yellow report desired tracking performance \cite{eic_YR}, which are highlighted as the brown dashed lines. In addition to the MAPS based silicon vertex and tracking detector, the EIC detector reference design \cite{ecce} also contains the AC coupled Low Gain Avalanche Diode (AC-LGAD) based outer tracker. The AC-LGAD outer tracker has fast timing resolution ($< 1~ns$), and can help suppressing backgrounds from neighboring collisions. A series of high precision forward heavy flavor hadron and jet measurements \cite{lanl_hf1, lanl_hf2, lanl_hf3} can be realized with the help of the proposed FST especially in constraining the heavy quark fragmentation function uncertainties in the high hadron momentum fraction region.

\section{Silicon Detector R$\&$D for the EIC}
\begin{figure}[!ht]
\centering
\includegraphics[width=0.85\textwidth]{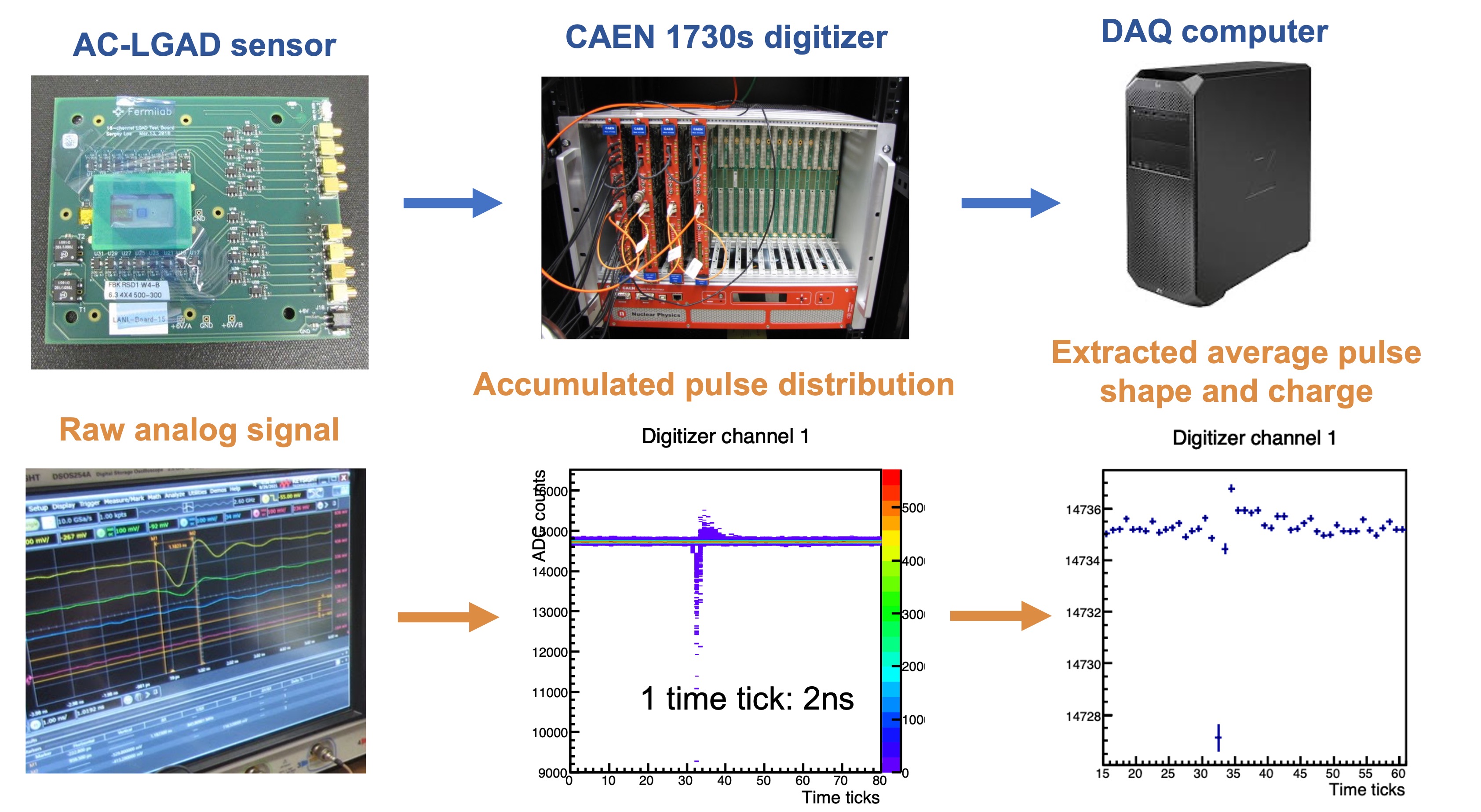}
\caption{The data processing and digitization scheme for AC-LGAD prototype sensor $^{90}Sr$ source tests. The readout chain starts with the analog outputs from the AC-LGAD sensor (top left), performs signal digitization with the CAEN 1730s digitizer (top middle) and carries out monitoring and final data processing in the DAQ computer (top right). The analog and digital data outputs for each step is shown in the bottom row.}
\label{fig:aclgad_bench}
\end{figure}

\begin{figure}[!ht]
\centering
\includegraphics[width=0.85\textwidth]{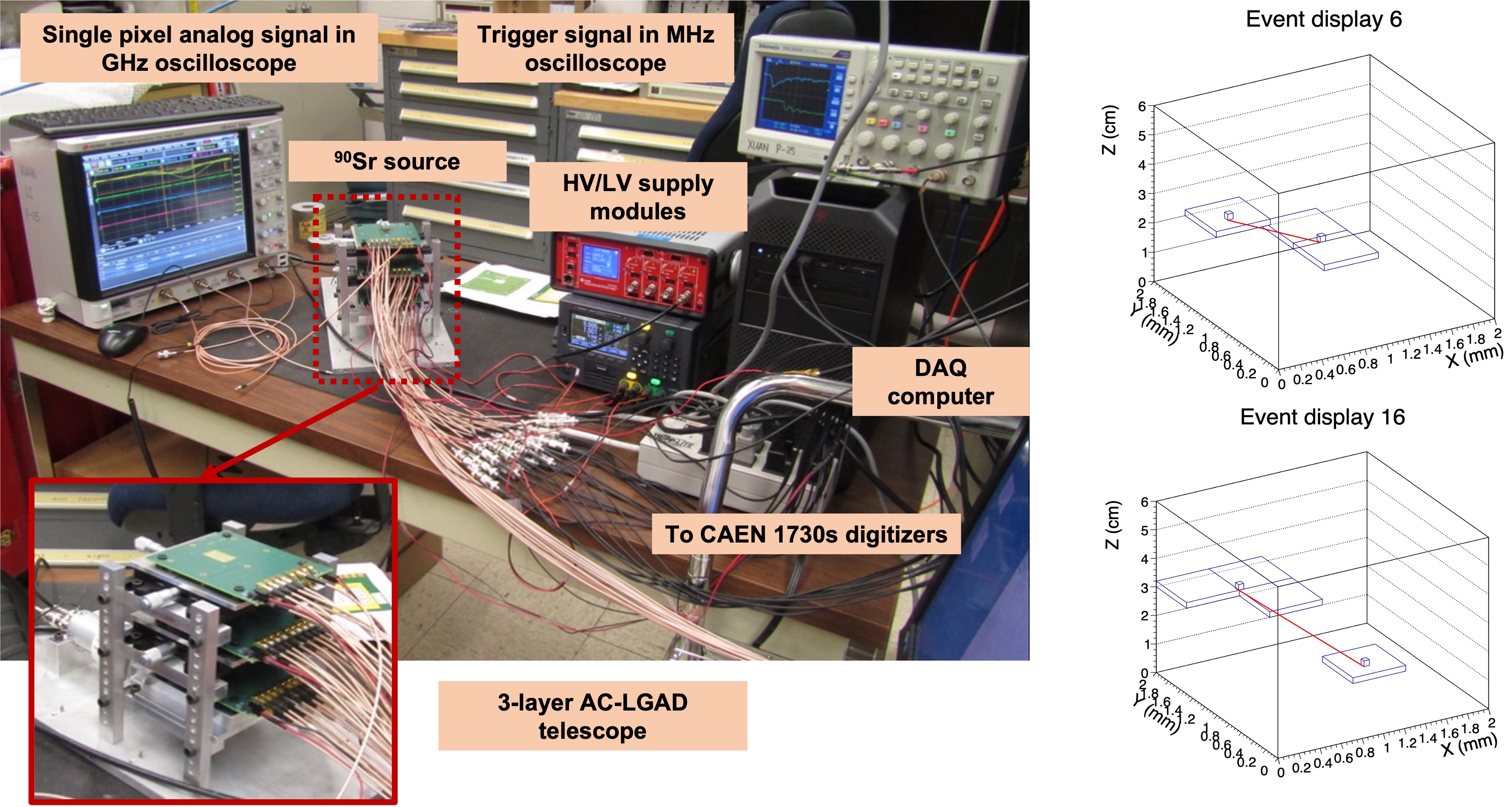}
\caption{The configuration of the 3-layer AC-LGAD telescope $^{90}Sr$ source tests is shown in the left and two event displays of tracks reconstructed with this telescope in $^{90}Sr$ source tests are shown in the right.}
\label{fig:aclgad_telescope}
\end{figure}

Prototype sensors of two advanced silicon technologies, AC-LGAD \cite{ac-lgad} and Depleted MAPS (DMAPS) (i.e., MALTA \cite{malta}) have been characterized with bench tests at LANL. The AC-LGAD prototype sensor consists of a $4 \times 4$ matrix of 500 $\mu m$ by 500 $\mu m$ pixels. This pixel matrix has been further divided into four $2 \times 2$ pixel groups. Charges can be shared by pixels belong to the same group and this feature can significantly improve the spatial resolution of AC-LGAD sensors \cite{ac-lgad}. A $^{90}Sr$ source has been used to characterize the AC-LGAD sensor performance. Figure~\ref{fig:aclgad_bench} shows the data processing scheme of the AC-LGAD sensor $^{90}Sr$ source bench tests, which utilizes the CAEN 1730s digitizer to perform the signal digitization. If all pixels on the AC-LGAD sensor are alive and have comparable amplitudes with each other, the AC-LGAD sensor is marked as good. A 3-layer AC-LGAD telescope has been assembled with three good AC-LGAD sensors. Figure~\ref{fig:aclgad_telescope} shows the configuration of the three-layer AC-LGAD sensor telescope $^{90}Sr$ tests, which utilizes the readout chain illustrated in Figure~\ref{fig:aclgad_bench}. A fast scintillator has been put underneath the bottom AC-LGAD sensor to provide the readout trigger. Track reconstruction has been performed from clustered pixels from different layers of this this AC-LGAD telescope in $^{90}Sr$ source tests. The right panel of Figure~\ref{fig:aclgad_telescope} shows two event displays of reconstructed beta decay electron tracks. The hit occupancy and tracking efficiency is under study with the AC-LGAD telescope. The tracking spatial and timing resolutions will be evaluated with the planned beam test facilities.

\begin{figure}[!ht]
\centering
\includegraphics[width=0.85\textwidth]{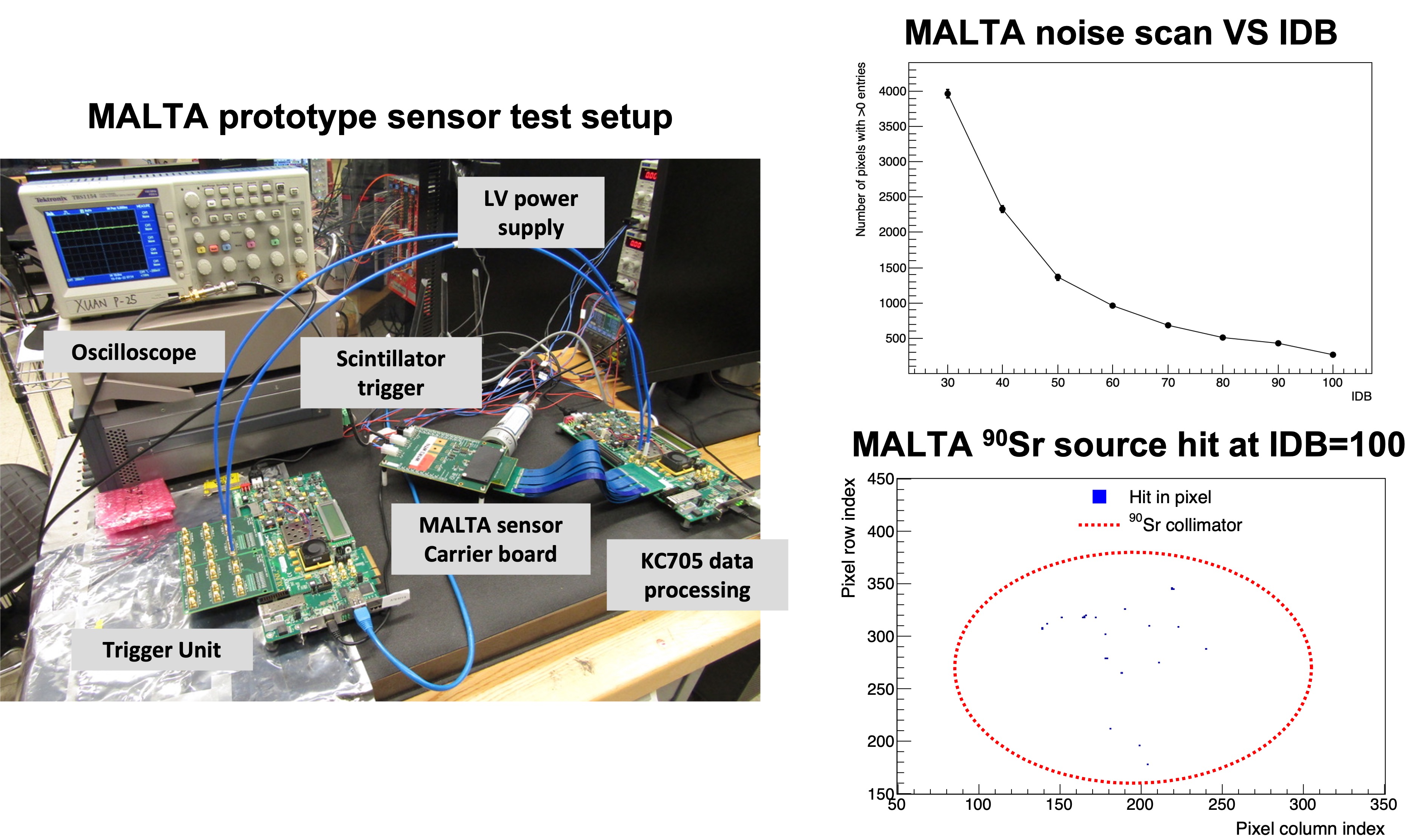}
\caption{The MALTA prototype sensor bench test configuration (left), the MALTA noise scan versus the IDB parameter (top right) and $^{90}Sr$ source hit occupancy in a MALTA prototype sensor with the IDB value at 100 (bottom right).}
\label{fig:malta_bench}
\end{figure}

A second test bench for MALTA prototype sensors has been set up as well. As shown in the left panel of Figure~\ref{fig:malta_bench}, the single MALTA sensor test setup consists of a KC705 XILINX based data processing module, a KC705 XILINX based trigger unit, a fast scintillator to provide the reference trigger, low power supplies for analog, digital and bias voltage operation. The MALTA prototype sensor contains 512 pixel columns, 512 pixel rows and the pixel size is 36.4 $\mu m$ by 36.4 $\mu m$ \cite{malta}. The MALTA sensor can inject charges onto individual pixels and the front-end threshold and noise can be determined by scanning the measured analog signals. A series of threshold and noise scan measurements have been carried out in different regions of the MALTA prototype sensor. The MALTA sensor can achieve better than 20 threshold over noise ratio and has good pixel performance uniformity over the active sensor area \cite{malta_tes}. 

The MALTA sensor threshold also depends on the Digital Analog Converter (DAC) parameters such as the discriminator bias current ("IDB") \cite{malta}. Optimization of the DAC setting has been carried out to achieve a good signal over noise ratio for the MALTA prototype sensor. The top right panel of Figure~\ref{fig:malta_bench} shows the IDB value dependent noise pixel rate from the MALTA sensor noise scan. The pixel noise rate reduces with increasing IDB values. After masking out high noisy rate pixels and selecting the IDB value at 100, a $^{90}Sr$ source has been placed on top of the MALTA sensor to study the pixel hit occupancy. The right bottom panel of Figure~\ref{fig:malta_bench} shows the MALTA pixel hit occupancy with the $^{90}Sr$ source. These hits are mostly located inside the source collimator coverage area and the region outside the source contains almost no hits. This result confirms that pixel noises have been successfully removed with the high IDB value and the noisy pixel masking. Further studies to check the physics hit cluster size to study the spatial resolution in triggered events are ongoing. The MALTA tracking spatial and temporal resolutions will be verified with CERN beam tests.

\section{Summary and Outlook}
The proposed Forward Silicon Tracker conceptual design, which has been included in the EIC project detector reference design \cite{ecce}, will enable high precision forward heavy flavor hadron and jet measurements at the future EIC. Promising results have been obtained from the lab bench tests for AC-LGAD and MALTA prototype sensors, which are candidate technologies for the EIC project detector tracking subsystem. Planned beam test studies will further validate the tracking spatial and timing resolutions, detector radiation hardness and readout chain integration for the EIC silicon vertex/tracking detector. Various parallel silicon detector R$\&$D efforts have been combined for the EIC project detector recently. The joint EIC project and general detector R$\&$D studies will provide valuable inputs for the EIC detector design optimization. An updated and integrated EIC vertex/tracking detector technical design will be delivered for the EIC CD2 approval and will serve as the guidance for the EIC detector construction, which is expected to start in 2025.

\end{document}